\documentclass[showpacs,preprintnumbers,amsmath,amssymb,superscriptaddress,aps,a4paper,prb,twocolumn]{revtex4}
\usepackage{graphicx}
\usepackage{dcolumn}
\usepackage{bm}

\begin{document}

\preprint{APS/123-QED}

\title{Impact of classical forces and decoherence in multi-terminal Aharonov-Bohm networks}

\author{E. Strambini}
\email{e.strambini@sns.it} 
\affiliation{NEST, Scuola Normale Superiore and CNR-INFM, 
		Piazza dei Cavalieri, 7, I-56126 Pisa, Italy}

\author{V. Piazza}
\affiliation{NEST, Scuola Normale Superiore and CNR-INFM, 
		Piazza dei Cavalieri, 7, I-56126 Pisa, Italy}

\author{G. Biasiol}
\affiliation{Laboratorio Nazionale TASC CNR-INFM, I-34012 Trieste, Italy}

\author{ L. Sorba}
\affiliation{NEST, Scuola Normale Superiore and CNR-INFM, 
		Piazza dei Cavalieri, 7, I-56126 Pisa, Italy}
\affiliation{Laboratorio Nazionale TASC CNR-INFM, I-34012 Trieste, Italy}

\author{F. Beltram}
\affiliation{NEST, Scuola Normale Superiore and CNR-INFM, 
		Piazza dei Cavalieri, 7, I-56126 Pisa, Italy}

\date{\today}

\begin{abstract}

Multi-terminal Aharonov-Bohm (AB) rings are ideal building blocks for quantum networks (QNs) thanks to their ability to map input states into controlled coherent superpositions of output states. We report on experiments performed on three-terminal GaAs/Al$_x$Ga$_{1-x}$As AB devices and compare our results with a scattering-matrix model including Lorentz forces and decoherence. Our devices were studied as a function of external magnetic field ($B$) and gate voltage at temperatures down to 350 mK. The total output current from two terminals while applying a small bias to the third lead was found to be symmetric with respect to $B$ with AB oscillations showing abrupt phase jumps between 0 and $\pi$ at different values of gate voltage and at low magnetic fields, reminiscent of the phase-rigidity constraint due to Onsager-Casimir relations. Individual outputs show quasi-linear dependence of the oscillation phase on the external electric field. We emphasize that a simple scattering-matrix approach can not model the observed behavior and propose an improved description that can fully describe the observed phenomena.
Furthermore, we shall show that our model can be successfully exploited to determine the range of experimental parameters that guarantee a minimum oscillation visibility, given the geometry and coherence length of a QN.
\end{abstract}


\maketitle

\section{Introduction}

A quantum network (QN) can be implemented by means of a set of nodes connected by electron waveguides that allow the coherent control of electron wavefunctions by means of external magnetic or electric fields.
One of the simplest QNs based on coherent electrons is a three-terminal Aharonov-Bohm (AB) ring with one input and two output channels \cite{3Rama2002}. When the linear superposition of the two output states can be tuned by means of an external field, this device realizes a qubit.
The implementation of single-qubit logic functions was proposed based on appropriately tailored multiterminal-terminal rings \cite{3Rama2002,4Rama2002}, making these QNs promising building blocks for quantum-computation architectures. Devices were theoretically analyzed by means of scattering-matrix approaches in the Landauer-B\"uttiker framework \cite{Datta1987,Datta1988}, free-electron-like node equations \cite{WuMahler1991,3Rama2002,4Rama2002}, and displaced Gaussian wavefunctions~\cite{Peeters2005}. The effect of Lorentz forces in three-terminal rings was theoretically studied in Ref.~6 by solving the time-dependent Schr\"{o}dinger equation. Despite this intense research activity, to the best of our knowledge these approaches have not yet been compared against experimental results on real devices.

In this work we shall extend existing theories based on scattering-matrix approaches by including the effect of decoherence and classical (Lorentz) forces in the description of the system and carry out this comparison by studying the low-temperature the coherent-transport properties of three-terminal AB rings. Our analysis shows that
the inclusion of these effects is necessary to fully understand the details of the observed behavior. Additionally, we shall identify a set of criteria that a device must satisfy to implement a functional QN and provide the operational range that guarantees a chosen value for oscillation visibility for a given set of device parameters and electronic coherence length.

Our devices were fabricated starting from a GaAs/Al$_x$Ga$_{1-x}$As heterostructure containing a high-mobility two-dimensional electron gas (2DEG). We observed AB oscillations in both output channels and in the total output of the system as a function of an external perpendicular magnetic field $B$. Oscillations show a non-trivial phenomenology as a function of gate voltage $V_{g}$: they evolve in a continuous way from a phase-rigid regime \cite{Casimir1945,Yacoby1995,Yacoby1996}, where the phase of the oscillations as a function of $B$ shows abrupt changes from 0 to $\pi$ when an external electric field $E$ is applied, to a linear dependence of the oscillation phase with $E$.

\section{Experimental results}

Three-terminal AB rings were fabricated from a two-dimensional electron gas (2DEG) confined 90 nm below the surface of a modulation-doped GaAs/Al$_x$Ga$_{1-x}$As heterostructure. At a temperature $T = 4$~K the unpatterned 2DEG density and mobility were found to be $2.1 \times{10^{11}}$~cm$^{-2}$ and $1.7\times{10^{6}}$~cm$^{2}$/Vs, respectively.

The ring geometry was defined by shallow plasma etching. The same processing step realized a set of lateral gates (labeled G$_1$ through G$_5$) that provide control over the electron density in each ring arm. A SEM image in artificial colors of one of our devices is shown in Fig.~1b. The ring external (internal) radius is 220 nm (90 nm).
Standard Ni/AuGe/Ni/Au (5 nm/180 nm/5 nm/100 nm) n-type Ohmic contacts (not shown in the figure) were fabricated to allow electrical access to the 2DEG.

The AB interference pattern can be tuned by exploiting the magnetic and/or electric AB effect \cite{Aharonov1959} by means of a perpendicular magnetic field or by changing the voltages $V_{G1}$, $V_{G2}$, and $V_{G3}$ applied to gates G$_1$, G$_2$, and G$_3$ respectively.

Measurements were performed in a $^{3}$He cryostat at 350 mK in a three-terminal configuration, where the injection contact was biased with an ac excitation signal $V_{ex} = 30$ $\mu$V at frequency of 170 Hz and currents $I_2$ and $I_3$ were measured respectively at leads $L_2$ and $L_3$ by means of current preamplifiers and phase-sensitive lock-in techniques. Blocking capacitors were present at the inputs of the preamplifiers to remove any unwanted dc component of the bias. In the following we shall refer to the ratio $g_i=I_i/V_{ex}$ as the conductance of the output lead $L_i$, with $i=2,3$.
Three nominally identical devices were investigated in detail and found to display the same behavior. In the following we shall focus on one of them, discussing when relevant the differences or the similarities observed in the others.

\begin{figure}[ht!]

\includegraphics[width=8.0cm]{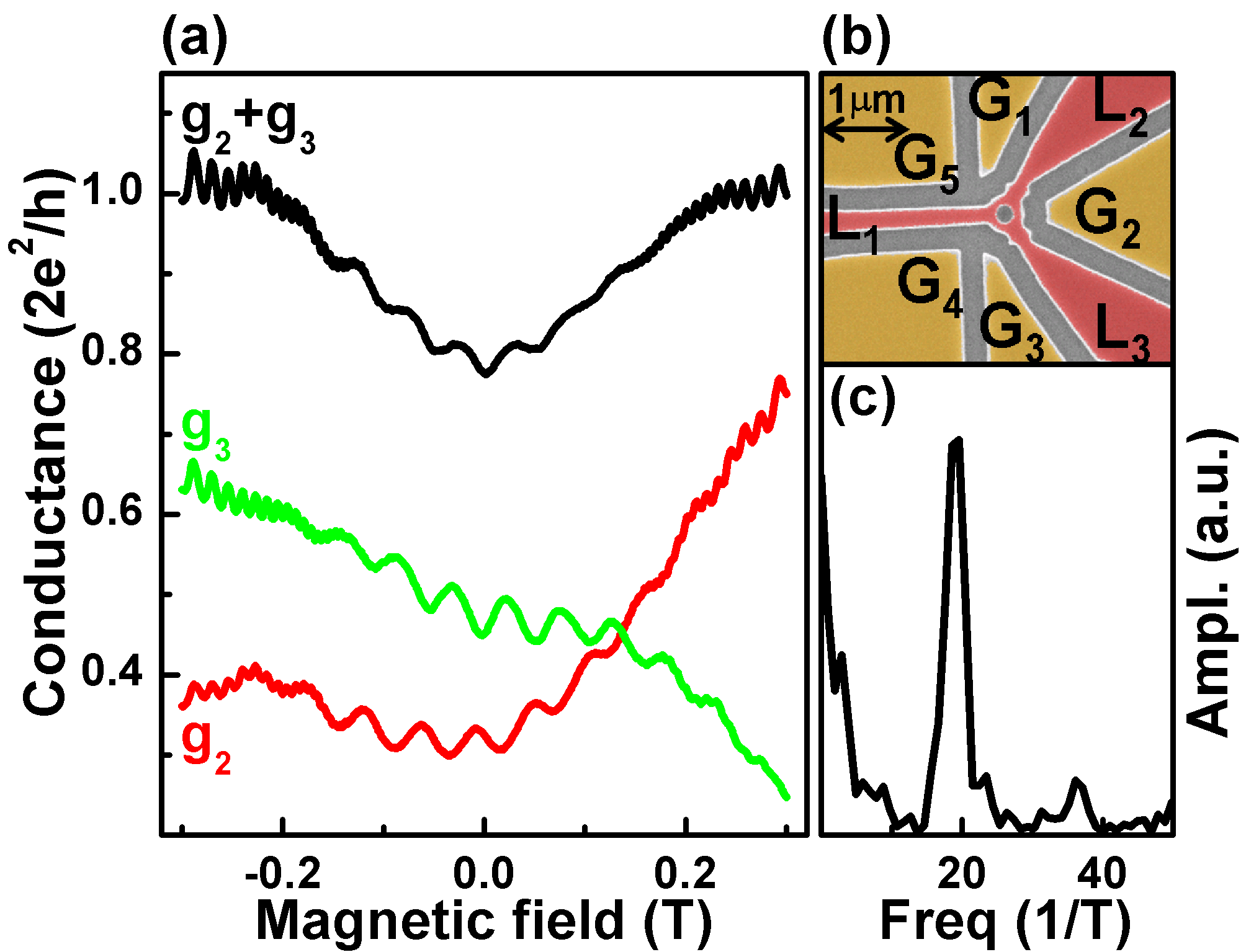}

\caption{(Color) (a) AB oscillations measured at a fixed gate voltages ($V_{G1} = \ldots = V_{G5} = 0.12 {\rm V}$) at 350 mK. The red and green lines represent the two output conductance, $g_2$ and $g_3$, while the black line shows the total conductance $g_2 + g_3$. (b) SEM image of the ring in artificial colors. The gray area corresponds to the 130-nm deep etching. The yellow areas represent the five gates (G$_1$ through G$_5$) while the central red region is the three-terminal AB ring with one input (L$_1$) and two output (L$_2$ and L$_3$) leads. The segment in the figure represents a length of 1~$\mu$m. (c) Fourier transform of the derivative with respect to $B$ of the total output current shown in Panel a.}
\label{fig1}
\end{figure}

Figure~1a shows $g_{2}$, $g_{3}$, and the total output conductance $g_{t} = g_{2} + g_{3}$ as a function of the perpendicular magnetic field at fixed gate voltages. AB oscillations are visible in both $g_{2}$ and $g_{3}$ and in their sum with a maximum visibility of approximately 0.2 indicating coherent transport across the ring.
The period of the oscillations, calculated by Fourier transforming the data (Fig.~1c), is $\sim$ 52.6 mT, corresponding to $h/e$ AB oscillations \cite{Aharonov1959}, where $e$ is the electron charge, and $h$ is the Plank constant, for a ring with an effective radius of 160 nm, in good agreement with the sample geometry. Similar values (155 nm and 170 nm) were found for the other devices studied.
The oscillating part of $g_{2}$ ($g_{3}$) is superimposed on a background that increases (decreases) as a function of $B$ and its amplitude decreases at higher (in modulus) magnetic fields. This behavior originates from an imbalance in the branching probability due to the Lorentz force \cite{Peeters2005} and leads to a suppression of the visibility of the oscillations that may limit the operation of a QN and should be taken into account in the design of the device.

The total output conductance was found to be symmetric with respect to the magnetic field as shown by the black curve in Fig.~1a. The second derivative of $g_{t}$ with respect to $B$ is reported in Fig.~2 as a function of $B$ and $V_{G2}$. It does display the same symmetry in the entire range of gate voltages explored. In the region between -0.1 T and 0.1 T of Fig.~2 abrupt jumps of the oscillation phase from 0 to $\pi$ can be seen at $V_{G2} \sim 0.17$ V and $V_{G2} \sim 0.125$ V, reminiscent of the phase-rigidity phenomena observed in two-lead, closed rings \cite{Yacoby1996,Aharony2006}.
As the modulus of the magnetic field approaches 0.2 T, these phase jumps become smoother, and evolve towards an almost continuous shift of the phase with gate bias. It is interesting to note that, within this continuous-evolution range, the oscillations at $B > 0$ present a phase shift dependence on the gate voltage that has an opposite sign with respect to those at $B < 0$: the oscillation maxima, shown as red areas in Fig.~2, drift to higher (in modulus) magnetic fields as the gate voltage is increased. We stress that this is not due to a change in the frequency of the oscillations, in fact the distance between neighboring maxima remains approximately constant.

\begin{figure}[ht!]
\includegraphics[width=8.0cm]{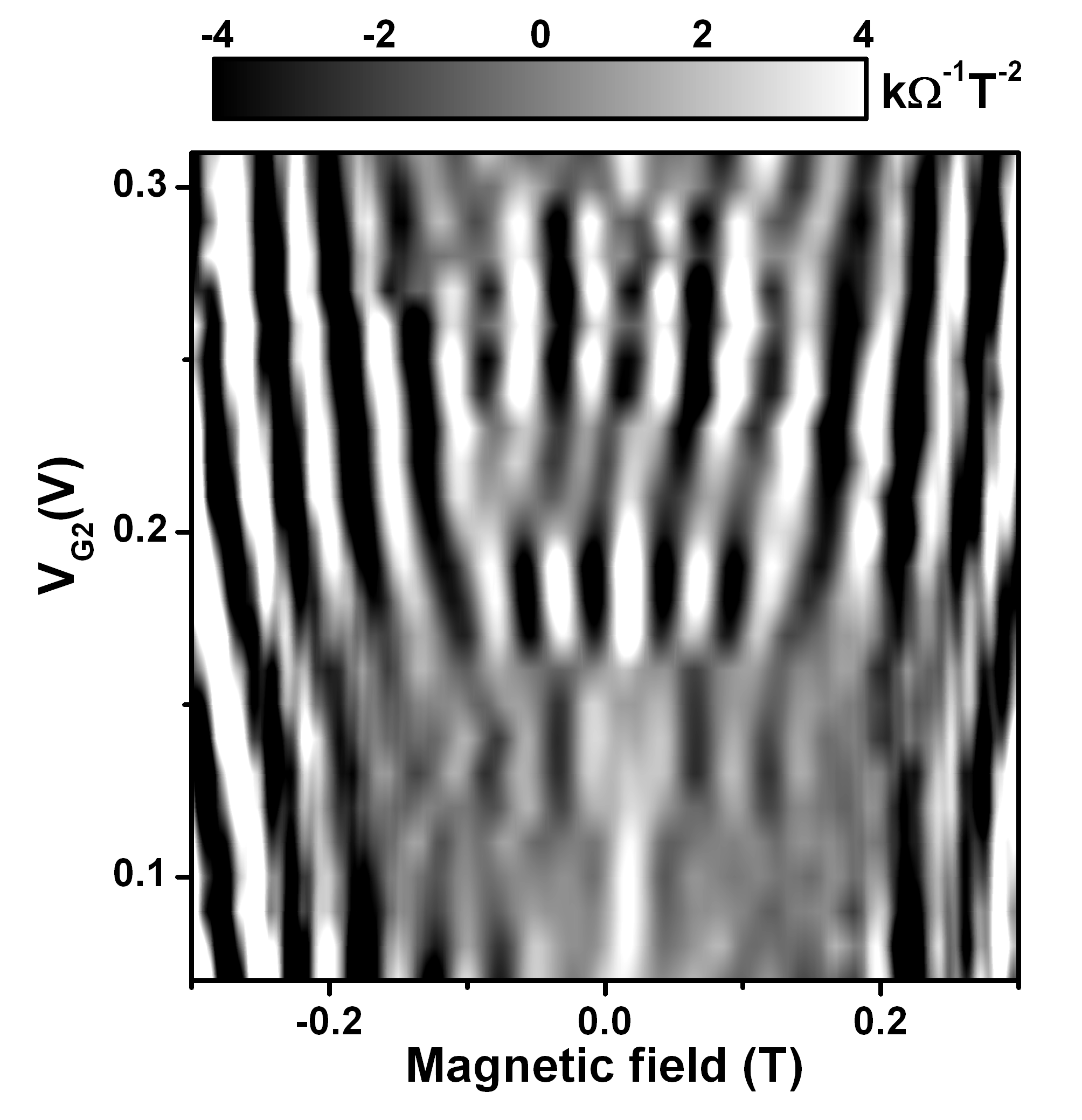}
\caption{Second-order derivative of the total conductance $g_2 + g_3$ with respect to the magnetic field plotted as a function of $B$ and $V_{G2}$.}
\label{fig2}
\end{figure}

A remarkably different behavior was observed in the evolution of the individual outputs as a function of $B$ and $V_{G2}$, shown in Fig.~3. In this case, the phase of the oscillations of $g_2$ evolves almost linearly with $V_{G2}$ in the entire range of magnetic fields and gate voltages explored. $g_3$ shows a similar behavior, but with an opposite dependence of the phase evolution on $V_{G2}$. The same behavior was observed in all the devices studied: the total output conductance displayed a transition from a phase-rigid regime to a linear regime at $|B| \approx 0.2$ T, while the individual outputs displayed a linear evolution pattern in all the range of magnetic fields explored.

\begin{figure}[ht!]
\includegraphics[width=8.0cm]{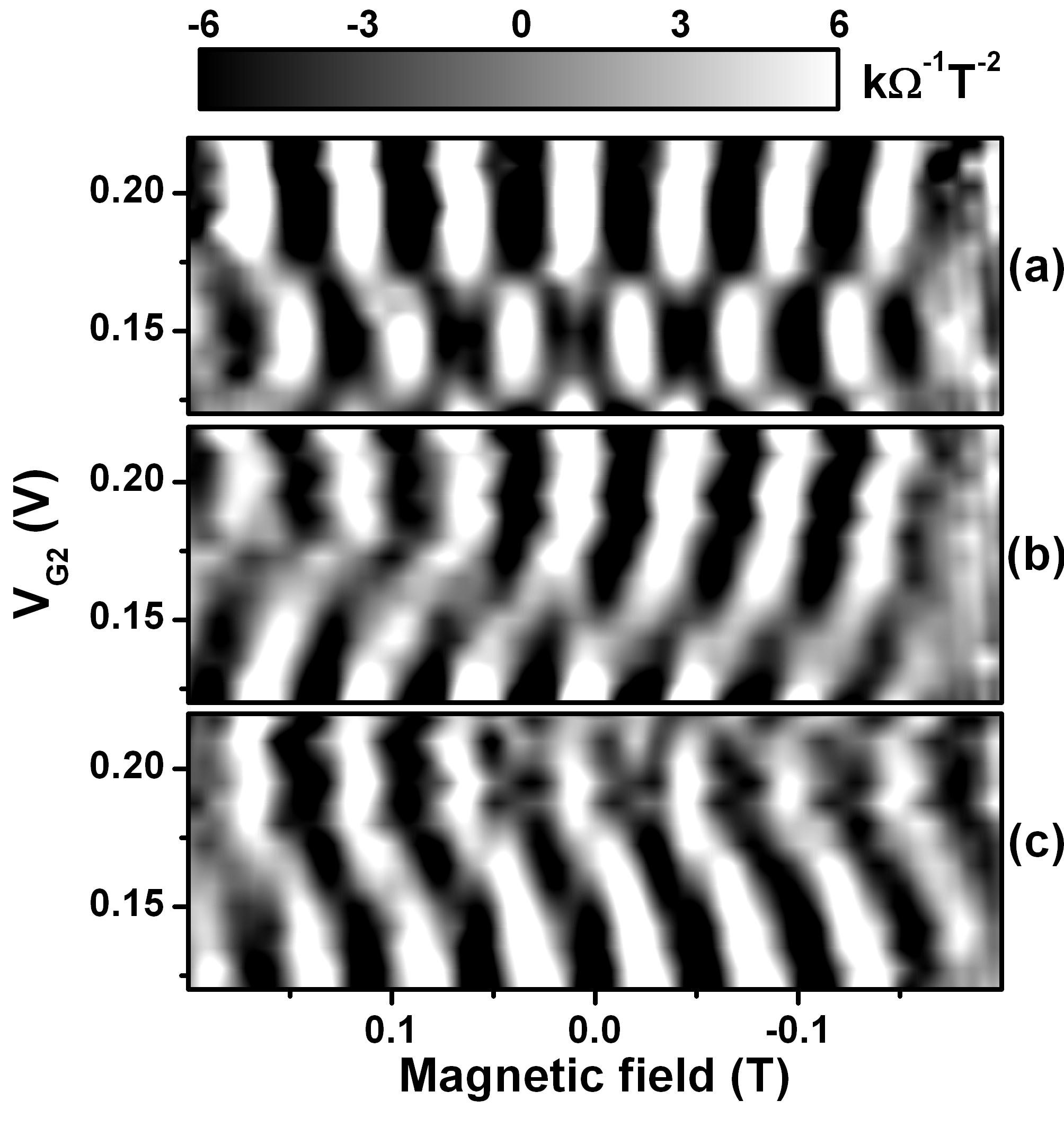}
\caption{Contour plots of the second-order derivatives of $g_2 + g_3$ (panel a), $g_2$ (panel b), and $g_3$ (panel c) with respect to the magnetic field as a function of $B$ and $V_{G2}$.}
\label{fig3}
\end{figure}

\section{Theory and simulations}

Here we shall propose a novel scattering-matrix approach that is able to explain in detail the observations reported in the previous section and will compare our numerical results with the experimental data. Following Refs.~3,4, electronic transport can be described in terms of transmission and reflection coefficients of a $N \times M$ scattering matrix $\mathbf{S}$ that in the most general case links $M$ input electron amplitudes ($\mathbf{u}$) with N output ones ($\mathbf{u'}$).
$$
\mathbf{u'}= \mathbf{S}\mathbf{u}.
$$

In the case of our three-lead devices, the propagation of an electron with a given energy is modeled by a $3\times 3$ complex matrix, $\mathbf{S}$, whose coefficients can be calculated given the scattering matrices of the three identical blocks shown in Fig.~4a that compose the ring.
Each individual block (Fig.~4b) is composed of a scattering center (black triangle in the figure) coupled to three leads. One of the leads represents one of the three ring arms (black square) where electron phase evolution is driven by wavefunction propagation and electric and magnetic AB effects.
Lorentz forces are modeled by appropriately changing the branching probability of the scattering center as a function of magnetic field, while decoherence effects are introduced in our scheme by decreasing the amplitude of the coherent wavefunction of the transmitted electron.

\begin{figure}[ht!]
\includegraphics[width=8.0cm]{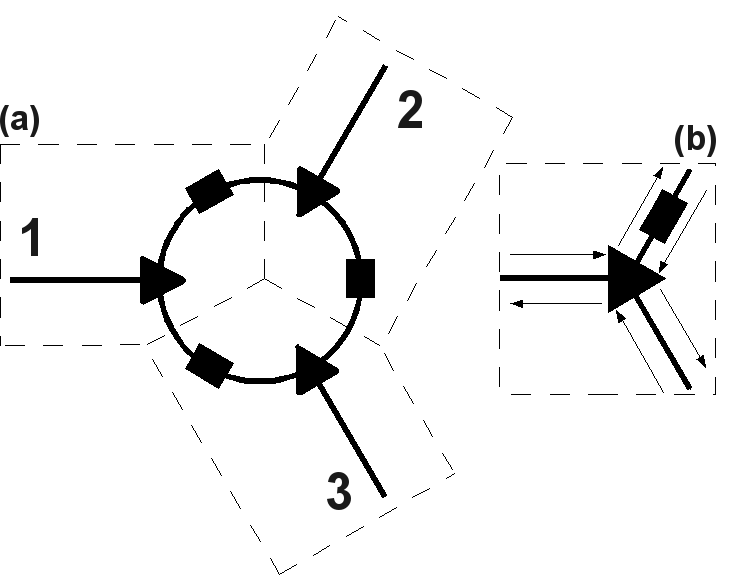}
\caption{(a) Complete scheme of the three-terminal AB ring. (b) Representation of a building block describing the coupling between one input or output lead and the ring.}
\label{fig4}
\end{figure}

The scattering center is represented by a $3\times 3$ scattering matrix. Current conservation requires this matrix to be unitary. We also restrict ourselves to the case of real matrix elements, which is equivalent to assume that no phase shifts are introduced by the scattering center.
Under this assumptions the matrix is orthogonal (O(3)) and can be represented as $\mathbf{\widetilde{M}} = \pm \mathbf{Q}_z(\alpha) \mathbf{Q}_x(\beta) \mathbf{Q}_z(\gamma)$, where the three independent parameters $\alpha$, $\beta$, $\gamma$ represent the three Euler's angles, $\mathbf{Q}_i(x)$ is a rotation of an angle $x$ around the $i$ axis, and the choice of the sign accounts for the two possible parity configurations.

This representation reproduces the symmetric branching described in Refs.~12,13 for $\alpha = 3 \pi /4$, $\gamma = \pi /4$. The parameter $\beta$ spans from $0$ to $\pi /2$ and represents the coupling parameter between input and output leads.

We assume that the effect of the Lorentz force on transmission and reflection probabilities of the scattering center in Fig.~4b is invariant under cyclic exchange of the leads. This symmetry of the system restricts the number of free parameters from three to one. Taking $\gamma$ as the independent variable, the other two angles can be written as:
$$\alpha = \pi /2 + \gamma $$
$$\beta = \arccos \left(-\frac{\sin 2\gamma}{2 + \sin 2 \gamma}\right).$$
By substitution, we obtain the following scattering matrix:
$$\mathbf{M} = \left(
\begin{array}{ccc}
a				&	b \sin \gamma	&	b \cos \gamma	\\
b \cos \gamma	&	a				&	b \sin \gamma	\\
b \sin \gamma	&	b \cos \gamma	&	a
\end{array}\right),
$$
where the coefficients $a$ and $b$ are given by:
$$a_{\pm} =\mp \frac{\sin 2 \gamma}{2 + \sin 2 \gamma}$$
$$b_{\pm} =\pm \sqrt{1-a^{2}}.$$

We select solutions $a=a_{+}$ and $b=b_{+}$ because they represent the realistic condition where transmitted electrons do not gain an extra $\pi$ phase.
As can be easily seen by examining $M$, the special cases $\gamma = 0$ and $\gamma = \pi /2$ describe the situation where an incoming electron is totally transmitted to one of the outputs or the other. Intermediate cases fall in the range $0 < \gamma < \pi /2$, the symmetric branching condition at zero magnetic field being obtained for $\gamma = \pi /4$.

As a first order approximation, we assume a linear dependence of the parameter $\gamma$ on the magnetic field:
$$\gamma = \frac{\pi}{4} \left(1 + \frac{B}{B_M}\right).$$ 

This assumption restricts the allowed values of the magnetic field within the range $-B_M \leq B \leq B_M$, where $B_M$ represents the magnetic field at which incoming electrons are fully deflected into only one of the two outputs.

Electron-phase evolution in the ring arms, including electric and magnetic AB phases and decoherence effects, is schematized by the black boxes of Fig.~4 and is introduced in our model by means of $2 \times 2$ scattering matrices:
$$
\left(
\begin{array}{cc}
0 				&	e^{i\phi /3 - i\theta - \xi} \\
e^{-i\phi /3 - i\theta - \xi}	&	0
\end{array}
\right),
$$
here $\phi = 2 \pi \Phi / \Phi_0$ ($\Phi_0$ is the AB quantum of flux $h/e$ and $\Phi$ the magnetic flux through the ring) represents the contribution to the electron-phase evolution due to the magnetic AB effect, $\xi = l / \lambda_c$ the ratio between the length of one arm of the ring ($l = 2 \pi r/3$) and the coherence length ($\lambda_c$), and $\theta$ the component due to electron propagation along the arm, which amounts to $k_f l$ ($k_f$ is the Fermi wavenumber), and the electric AB effect. In our scheme $\xi$ models decoherence phenomena occurring in our system by decreasing the wavefunction amplitude as the electron propagates along the arm.
This allows to empirically introduce the suppression of the oscillation amplitude observed in presence of decoherence at the output of QNs in a computationally more efficient way compared to the conventional approach consisting in introducing random phase shifts with a given probability distribution and averaging the output signal over it.

The scattering matrix of the $n$-th block shown in Fig.~4b ($n = 1, \ldots, 3$) composing the ring, including the scattering center and the phase evolution due to the propagation along the arm, is obtained by the combination of the two scattering matrices described above:

\begin{widetext}
$$\mathbf{J_n}=\left(
\begin{array}{ccc}
M_{11}					&	M_{12}\, e^{-i\phi /3 - i\theta_n - \xi} 	&	M_{13}	\\
M_{21}\, e^{i\phi /3 - i\theta_n - \xi}	&	M_{22}\, e^{- 2 i\theta_n - 2\xi}		&	M_{23}\, e^{i\phi /3 - i\theta_n - \xi}\\
M_{31}					&	M_{32}\, e^{-i\phi /3 - i\theta_n - \xi}	&	M_{33}
\end{array}\right).
$$
\end{widetext}

We introduced three different phases, $\theta_n$, to take into account that the electron propagation in each arm can be independently controlled by means of the three side gates: G$_1$, G$_2$, and G$_3$. In the case where more than one one-dimensional subband is available for transport, the contribution from each of them can be evaluated by calculating $\theta_n$ with the corresponding value of $k_f$. This would reflect in the total output current as a suppression of the AB oscillations, but would not give rise to a qualitative change in the oscillation behavior as a function of gate voltage and magnetic field as can be easily verified by calculating the transmission and reflection coefficients by choosing different values of the parameters $\theta_n$. In the following we shall restrict our description to the case of a single propagation mode.
Since in our experiments we keep the gates G$_1$ and G$_3$ at a constant value, we set $\theta_1 = \theta_3= \pi/2$. As argued above, a different choice would have led to a shift in the calculated characteristics of no impact in the comparison with the experiment. The remaining phase $\theta_2$ will be used to model the effect of gate G$_2$.
The scattering matrix $\mathbf{S}$ of the entire ring is calculated by combining three of these matrices, one per ring node.
The probabilities $T_2$ and $T_3$ for an electron to be transmitted to the two outputs of the system and $R$ to be reflected backwards into the input lead are obtained by calculating the squared modules of the components of $\mathbf{u'}$, given the input condition $\mathbf{u}=(1,0,0)$. This situation corresponds to injecting electrons in the ring only from the input (left) terminal. We wish to emphasize that, since in our model electrons that are scattered incoherently are removed from the system, the sum of the reflection and transmission probabilities does not need to equal one.

The impact of decoherence and Lorentz forces on the behavior of a three-terminal ring is highlighted in Fig.~5, which reports our results for a ring with a radius of $160\,{\rm nm}$:
Panel (a) shows the calculated total output probability without these two effects, Panel (b) shows the results of the calculations for $B_M = 370\,{\rm mT}$, $\lambda_c = 320\,{\rm nm}$. These values were found by means of a best fitting procedure to the experimental data.

\begin{figure}[ht!]
\includegraphics[width=8.6cm]{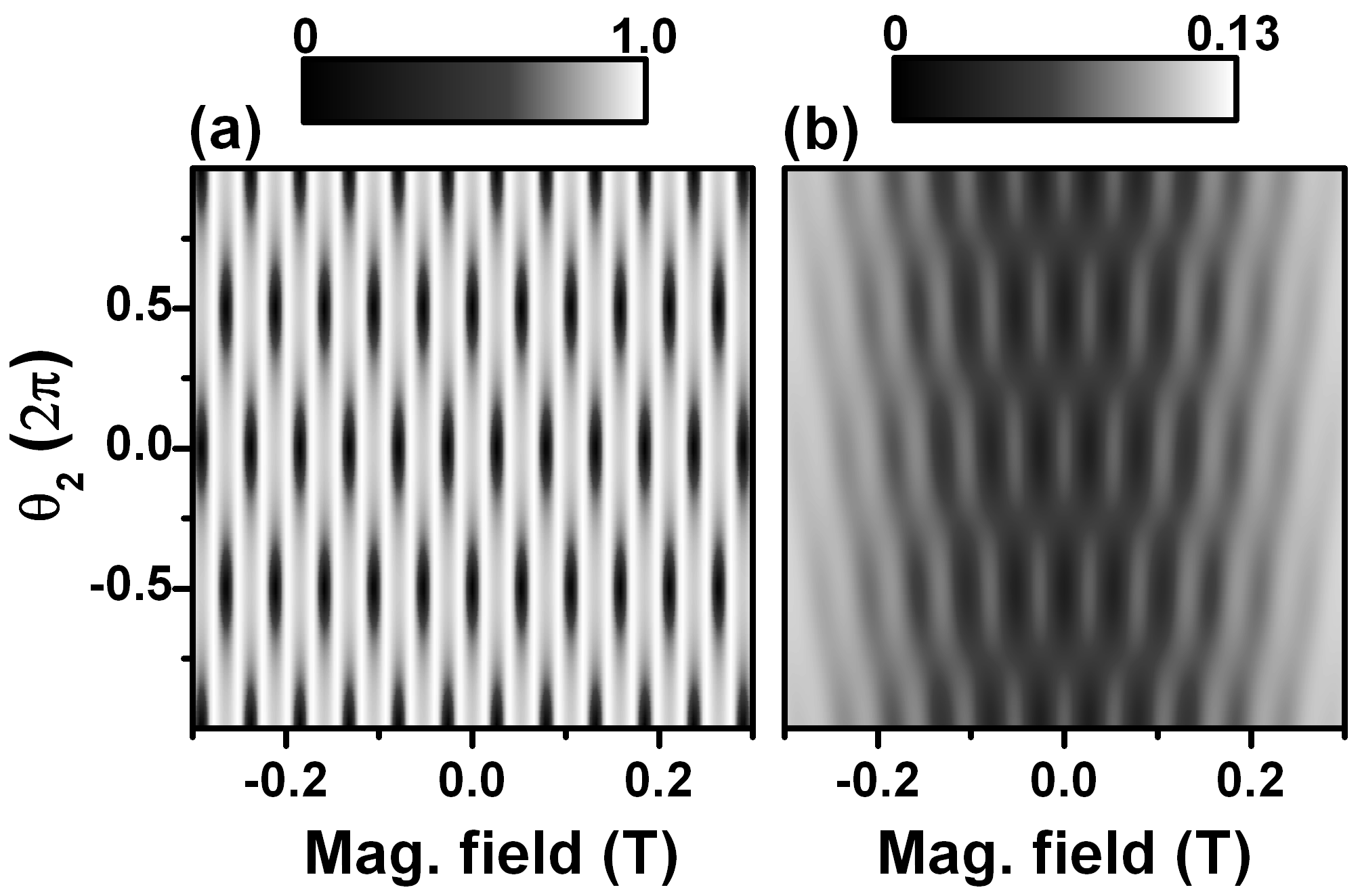}
\caption{Contour plots of the calculated total transmission probability as a function of magnetic field and electric phase shift (${\theta}$). (a) Conventional scattering-matrix approach, with no Lorentz forces and decoherence ($1/B_M = 0$, $1/\lambda_c = 0$), (b) Modified scattering-matrix calculations including Lorentz force ($B_M = 370\,{\rm mT}$) and decoherence ($\lambda_c = 320\,{\rm nm}$).}
\label{fig5}
\end{figure}

In Fig.~5a a strong phase-rigidity pattern dominates the entire range of external magnetic and electric fields, as expected for perfectly coherent, closed systems.
On the other hand, in Fig.~5b a region reminiscent of phase rigidity is present only in a low-magnetic-field region ($|B|\lesssim B_M/3$). At higher field values the oscillation phase evolves linearly with the absolute value of $B$ and the oscillation amplitude is suppressed due to the Lorentz force, as discussed above. It is clear from the comparison of the two figures that accounting for the Lorentz force and decoherence drastically changes the behavior of the total conductance as a function of the external magnetic and electric fields.

In order to allow the comparison of our model results to the experimental data, we show in Fig.~6 $d^2 T_2/dB^2$, $d^2 T_3/dB^2$, and their sum, calculated for the same parameters of Fig.~5b, as a function of magnetic field and gate voltage. The phase of the oscillations reported in Figs.~6b and 6c, shows a linear -- but opposite in sign -- dependence from the gate voltage, as observed in our devices.

\begin{figure}[ht!]
\includegraphics[width=8.0cm]{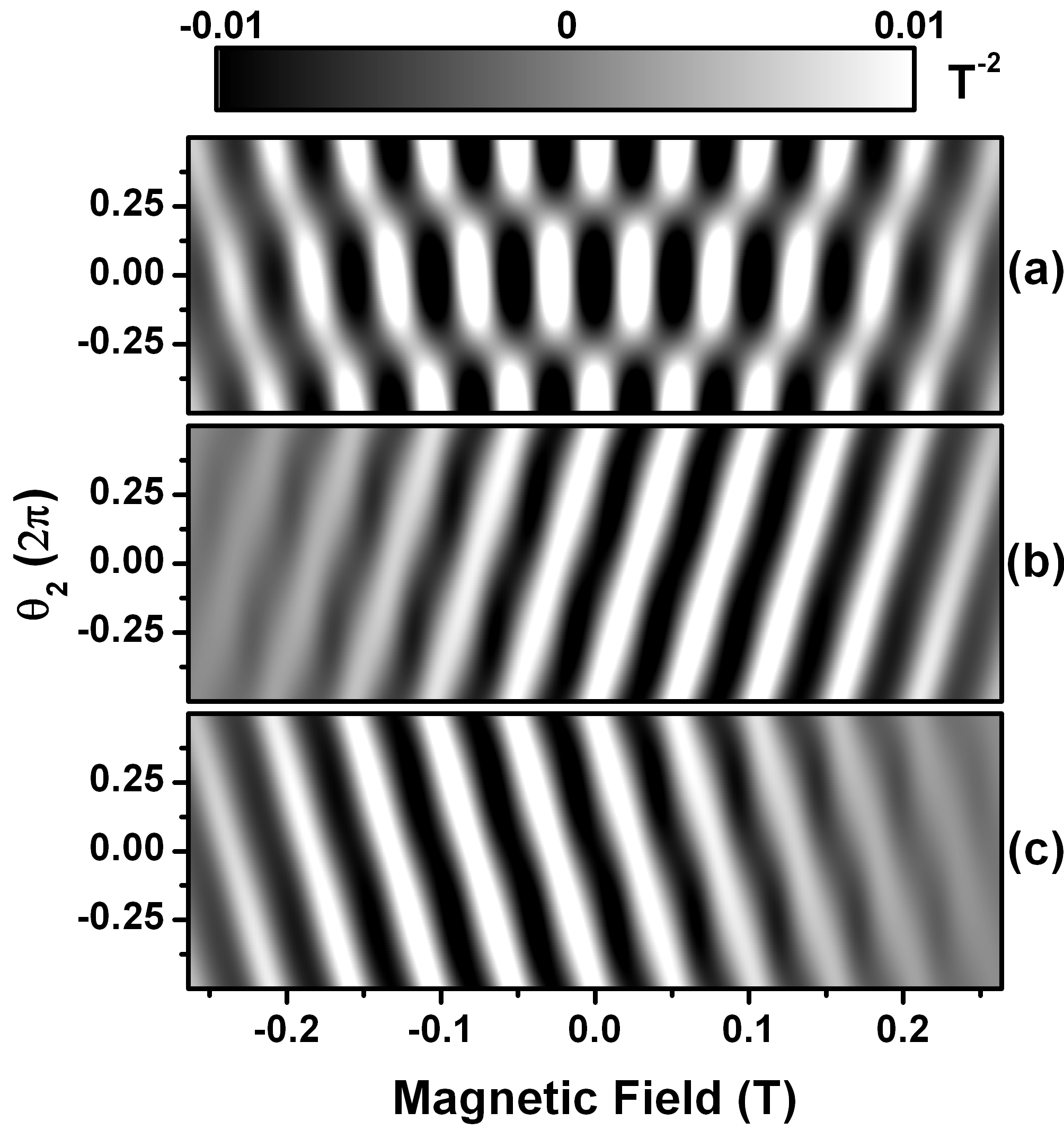}
\caption{Contour plots of the second derivative of the calculated transmission probabilities as a function of magnetic-field and electric phase shift (${\theta_2}$). (a) Total transmission probability. (b),(c) Individual channel transmission probabilities.}
\label{fig6}
\end{figure}

A further comparison between simulated and experimental total output of our ring is plotted in Fig.~7. Inspection of this figure confirms the ability of our model to provide an accurate description of the experimental data.

\begin{figure}[ht!]
\includegraphics[width=8.0cm]{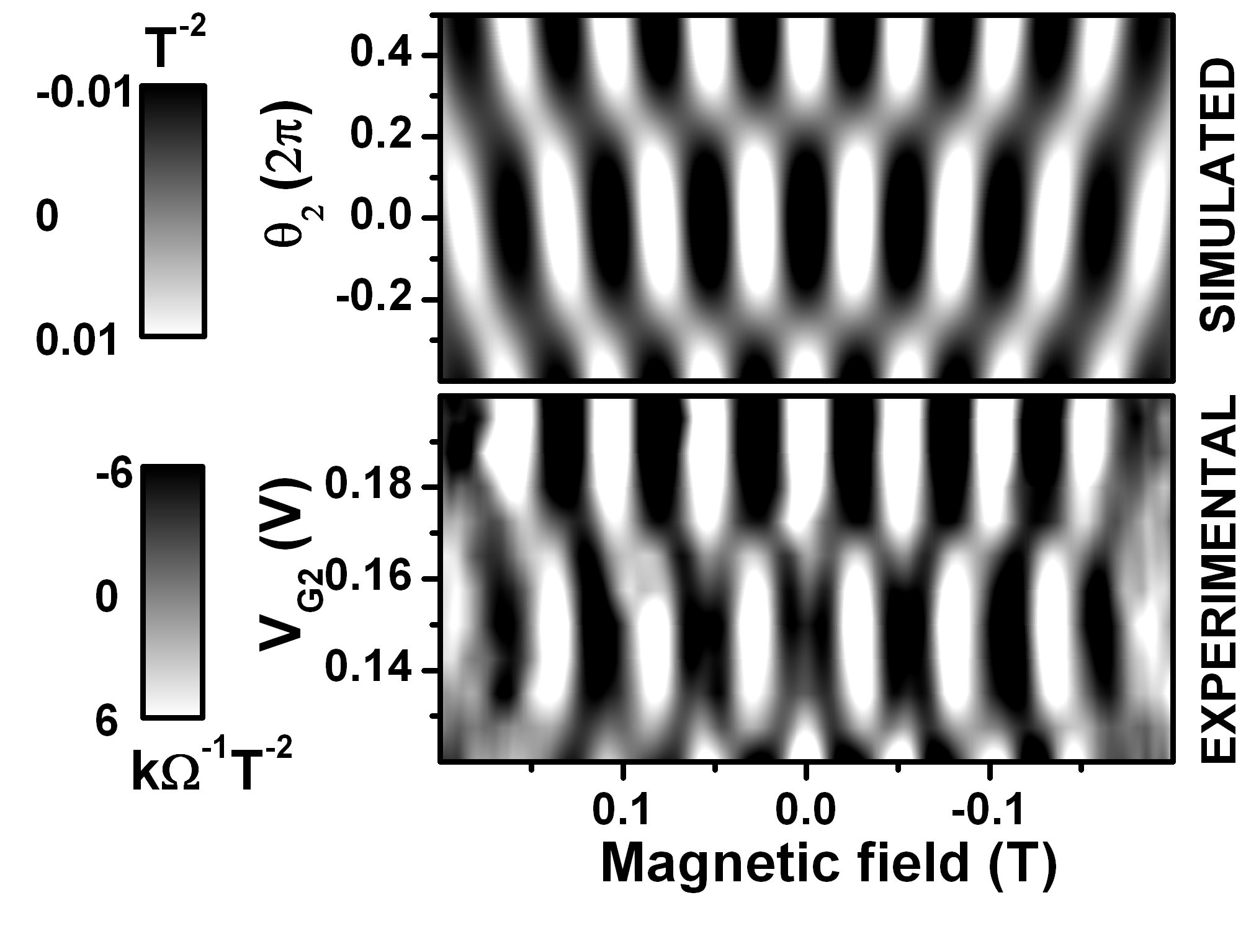}
\caption{Comparison between the simulated total conductance and the second-order derivative with respect to $B$ of the measured $g_2 + g_3$ as a function of the external magnetic and electric fields.}
\label{fig7}
\end{figure}

\section{Range of operation of a quantum network}

As discussed above, Lorentz force and decoherence limit the performance of a QN and should be taken into account when designing such a device. Our theoretical model can be exploited to quantify the impact of these effects on the operation of a QN. Indeed, by setting the minimum acceptable oscillation contrast for the operation of a given QN, our model allows one to identify a region in parameter space where the chosen requirements are met.
To this end, it is particularly convenient to fix the model variables that are related to external fields (parametrized here by $\phi$ and $\theta_2$) and use as independent, dimensionless variables $\xi$ and $\Phi_0 / (B_M S)$, where $S$ is the ring area, that are related to the sample characteristics. $B_M$ can be estimated from the effective arm width $W$, by assuming that almost complete steering of the electron in one of the two outputs occurs when $r_c \lesssim W$, $r_c^2 = h / (e B)$ being the cyclotron radius. This condition is equivalent to $B \gtrsim \Phi_0 / W^2$ and allows to calculate $B_M$ as: $B_M = \Phi_0 / W^2$.
In the case of the present sample the value of $B_M$ yielded by the fitting procedure ($370\,{\rm mT}$,) allows to estimate an effective width $W \approx 100\,{\rm nm}$, consistent with the value of the sample arm width (Fig.~1b). In the following we shall set the minimum visibility at 15\% to match the value of the visibility observed in our devices.

In our model the fraction of electrons that loose phase-coherence is removed from the system and can be calculated as:
$$1 - (R + T_2 + T_3).$$
Also the fraction of coherent electrons that are not useful for further computations, owing to the imbalance created by Lorentz force and decoherence, must be taken into account. This can be evaluated by taking the ratio between the amplitude of the AB oscillations and the background of the signal transmitted from each output. We evaluate the contribution of the two effects separately: each of them defines a subset of the $\xi$, $\Phi_0 / (B_M S)$ plane, shown respectively in Fig.~8a and Fig.~8b. The intersection of the two regions (shown in Fig.~8c and calculated at different values of $\Phi / \Phi_0$) corresponds to the set of experimental parameters that leads to correct operation of the QN. Figure~8 shows the operation region taking into account decoherence effects (Panel a) and imbalance (Panel b). As can be observed from Fig.~8c, requiring the operation of the QN within specifications at higher values of $\Phi / \Phi_0$, results in a more stringent demand on the sample parameters, owing to the oscillation visibility decreasing at increasing values of the magnetic field.
This approach can be straightforwardly extended to more complex QNs and yields the conditions for a given performance requirement. We believe it can be a useful tool for the design of practical QNs.

\begin{figure}[ht!]
\includegraphics[width=8.0cm]{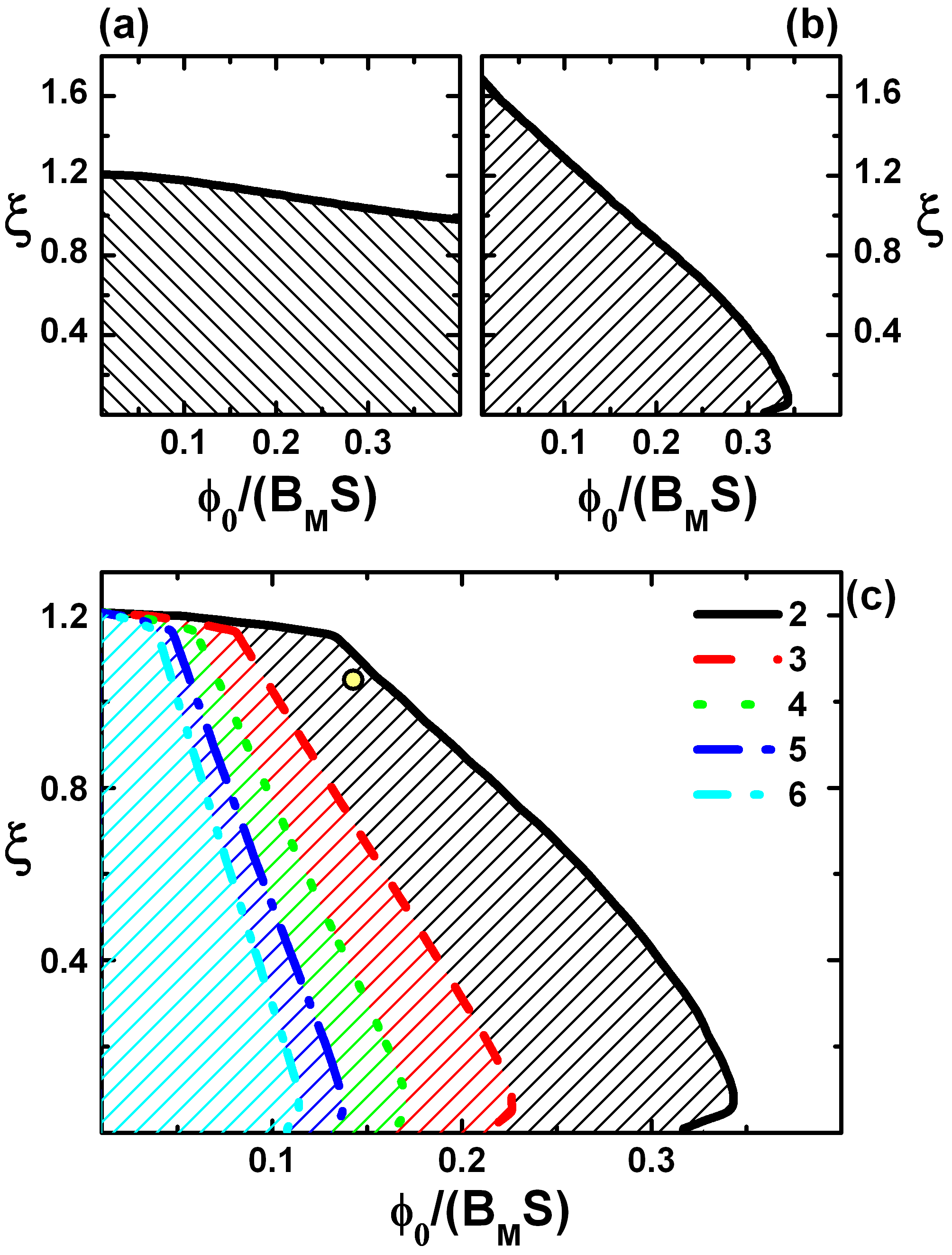}
\caption{ (Color)
(a) Subset of the plane identified by $\xi$, the ratio between the length of one arm of the ring and the coherence length, and by the ratio $\Phi_0 / (B_M S)$, which quantifies the impact of the Lorentz force, where the fraction of electrons that maintain phase coherence when traveling across the device is larger than 15\%, calculated at a magnetic field corresponding to $\Phi / \Phi_0 = 2$, and a gate voltage corresponding to $\theta = 0$.
(b) Subset of the $\xi$, $\Phi_0 / (B_M S)$ plane where the fraction of coherent electrons that contributes to the network operation is larger than 15\%.
(c) Subset of the $\xi$, $\Phi_0 / (B_M S)$ plane where the two conditions depicted in Panels a and b are satisfied at $\theta = 0$ and for different values of $\Phi / \Phi_0 $. The filled dot represents the values of the parameters $\xi$ and $\Phi_0 / (B_M S)$ that yielded the best fit to the experimental data.}
\label{fig8}
\end{figure}

\section{Conclusions}
Three-terminal ring devices were fabricated and studied at cryogenic temperatures showing clear AB oscillations as a function of external magnetic and electric fields. The oscillations of the total output signal were found to exhibit an evolution from a phase-rigid pattern to a linear dependence of the phase on the external electric field at increasing (in modulus) magnetic fields. The individual outputs displayed a linear dependence of the phase on the electric field. We have shown by means of a novel scattering-matrix approach that this peculiar behavior can be explained only by taking into account the competition between decoherence effects and classical (Lorentz) forces. Our approach can be successfully exploited to determine the range of experimental parameters that guarantee a given oscillation visibility, given the geometry and coherence length of a QN. We emphasize that this model can be easily applied to more complex QN systems while still remaining computationally efficient.

E.~S. wish to thank the Fondazione Silvio Tronchetti Provera for financial support during this work.


\begin{thebibliography}{}
\bibitem{3Rama2002}C.~H.~Wu and D.~Ramamurthy, Phys.~Rev.~B {\bf 65}, 075313 (2002);
\bibitem{4Rama2002}D.~Ramamurthy, and C.~H.~Wu, Phys.~Rev.~B {\bf 66}, 115307 (2002);
\bibitem{Datta1987}S.~Datta, M.~Cahay, and M.~McLennan, Phys.~Rev.~B {\bf 36}, 5655 (1987);
\bibitem{Datta1988}M.~Cahay, M.~McLennan, and S.~Datta, Phys.~Rev.~B {\bf 37}, 10125 (1988);
\bibitem{WuMahler1991}C.~H.~Wu, and G.~Mahler, Phys.~Rev.~B {bf 43}, 5012 (1991);
\bibitem{Peeters2005}B.~Szafran and F.~M.~Peeters, Europhys.~Lett.~{\bf 70}, 810 (2005);
\bibitem{Casimir1945}H.~B.~G.~Casimir, Rev.~Mod.~Phys.~{\bf 17}, 343 (1945);
\bibitem{Yacoby1995}A.~Yacoby, M.~Heiblum, D.~Mahalu, and H.~Shtrikman, Phys.~Rev.~Lett.~{\bf 74}, 4047 (1995);
\bibitem{Yacoby1996}A.~Yacoby, R.~Schuster, and M.~Heiblum, Phys.~Rev.~B {\bf 53}, 9583 (1996).
\bibitem{Aharonov1959}Y.~Aharonov and D.~Bohm, Phys.~Rev.~{\bf 115}, 485 (1959);
\bibitem{Aharony2006}A.~Aharony, O.~Entin-Wohlman, T.~Otsuka, S.~Katsumoto, H.~Aikawa, and K.~Kobayashi, Phys.~Rev.~B {\bf 73}, 195329 (2006);
\bibitem{Buttiker1985}M.~B\"uttiker, Phys.~Rev.~B {\bf 32}, 1846 (1985);
\bibitem{Buttiker1984}M.~B\"uttiker, Y.~Imry, and  M.~Ya.~Azbel, Phys.~Rev.~A {\bf 30}, 1982 (1984);
\end{thebibliography}
\end{document}